\documentclass[10pt,conference]{IEEEtran}
\usepackage{cite}
\usepackage{amsmath,amssymb,amsfonts}
\usepackage{graphicx}
\usepackage{textcomp}
\usepackage{xcolor}
\usepackage[hyphens]{url}

\def\BibTeX{{\rm B\kern-.05em{\sc i\kern-.025em b}\kern-.08em
    T\kern-.1667em\lower.7ex\hbox{E}\kern-.125emX}}

\newcommand{\ignore}[1]{}
\usepackage[pass]{geometry}
\usepackage{fancyhdr}
\usepackage[normalem]{ulem}
\usepackage{hyperref}
\usepackage{color}
\usepackage{soul}
\usepackage{algorithm}
\usepackage{algpseudocode} 

\usepackage{pbox}
\usepackage{booktabs}
\usepackage{microtype}
\usepackage{float}
\usepackage{tikz}
\usepackage{xspace}
\usepackage{enumitem}
\usepackage{subcaption}
\usepackage{booktabs}

\usepackage{tabularx}
\newcolumntype{P}{>{\raggedright\arraybackslash}X}
\newcolumntype{C}{>{\hsize=0.5\hsize}X}
\usepackage{makecell}
\usepackage{multirow}
\usepackage{pifont}

\usepackage{hyperref}
\usepackage{framed, multido}
\usepackage{xspace}

\everymath{\mathtt{\xdef\tmp{\fam\the\fam\relax}\aftergroup\tmp}}
\RequirePackage[varqu]{zi4} 

\usepackage{xparse}
\makeatletter
\NewDocumentCommand{\LeftComment}{s m}{%
  \Statex \IfBooleanF{#1}{\hspace*{\ALG@thistlm}}\(\triangleright\) #2}
\makeatother

\algnewcommand{\LeftCommentFirst}[2]{\Statex\hspace{#1} \(\triangleright\) #2}

\RequirePackage[T1]{fontenc} \RequirePackage[varqu]{zi4} \RequirePackage[libertine]{newtxmath}
\algdef{SE}[SUBALG]{Indent}{EndIndent}{}{\algorithmicend\ }%
\algtext*{Indent}
\algtext*{EndIndent}

\newcommand{\sys}[0]{\textsc{MoCA}\xspace}

\newcommand{\hw}[0]{\sys hardware\xspace}
\newcommand{\sw}[0]{\sys runtime system\xspace}
\newcommand{\compute}[0]{\texttt{compute}\xspace}
\newcommand{\mem}[0]{\texttt{mem}\xspace}

\title{MoCA: \underline{M}em\underline{o}ry-\underline{C}entric, \underline{A}daptive Execution \\for Multi-Tenant Deep Neural Networks}
\author{
\IEEEauthorblockN{Seah Kim, Hasan Genc, Vadim Vadimovich Nikiforov, \\Krste Asanovi{\'c}, Borivoje Nikoli{\'c}, Yakun Sophia Shao}
\IEEEauthorblockA{University of California, Berkeley\\
\texttt{\{seah,hngenc,vnikiforov,krste,bora,ysshao\}@berkeley.edu}
}
}

\begin{document}
\maketitle
\pagestyle{empty} 

\begin{abstract}
Driven by the wide adoption of deep neural networks (DNNs) across different application domains, multi-tenancy execution, where multiple DNNs are deployed simultaneously on the same hardware, has been proposed to satisfy the latency requirements of different applications while improving the overall system utilization. 
However, multi-tenancy execution could lead to undesired system-level resource contention, causing quality-of-service (QoS) degradation for latency-critical applications.

To address this challenge, we propose \sys\footnote{https://github.com/ucb-bar/MoCA}, an adaptive multi-tenancy system for DNN accelerators.
Unlike existing solutions that focus on compute resource partition, \sys dynamically manages shared memory resources of co-located applications to meet their QoS targets.
Specifically, \sys leverages the regularities in both DNN operators and accelerators to dynamically modulate memory access rates based on their latency targets and user-defined priorities so that co-located applications get the resources they demand without significantly starving their co-runners. 
We demonstrate that \sys improves the satisfaction rate of the service level agreement (SLA) up to 3.9$\times$ (1.8$\times$ average), system throughput by 2.3$\times$ (1.7$\times$ average), and fairness by 1.3$\times$ (1.2$\times$ average), compared to prior work.

\end{abstract}

\section{Introduction}\label{introduction}
{\rm

Recent advances in deep learning have led to the broad adoption of DNN-based algorithms in tasks such as object detection~\cite{resnet}, natural-language processing~\cite{bert}, AR/VR~\cite{illixr,edge-facebook,hda-hpca2021}, and robotics~\cite{mavbench-micro2018,dl-robotics}.
As a result, DNNs have become a core building block for a diverse set of applications running on devices ranging from edge platforms to data centers, many of which need to be executed at the same time to meet their latency requirements~\cite{mlperf-training, mlperf-inference-isca2020}.

To support the concurrent execution of these applications, \textit{multi-tenancy} execution, where multiple applications are colocated on the same hardware, is required to improve the overall efficiency of the system.
The key challenge of supporting effective multi-tenancy execution in DNN accelerators is the performance variability caused by contention for shared resources, especially for user-facing applications with strict latency requirements.
Recently, new techniques have been proposed to either temporally interleave the execution of multiple DNNs~\cite{layerweaver,AI-MT,PREMA} or spatially partition the compute resources of DNN accelerators~\cite{hda-hpca2021,Planaria,Veltair} to support multi-tenancy execution in DNN accelerators.

Although previously proposed microarchitecture or scheduling techniques are effective in partitioning the compute resources of accelerators, little attention has been paid to the management of shared memory subsystems, e.g., the shared system cache and DRAM, where multiple accelerators and general-purpose cores could compete for bandwidth and capacity.
The absence of system-level solutions to adaptive contention management would lead to unexpected end-to-end performance degradation.


To address this challenge, we present \sys, a memory-centric, adaptive accelerator architecture that supports efficient multi-tenancy execution for DNN workloads.
In contrast to prior work that focuses on the efficient sharing of accelerator compute resources, \sys aims to adaptively manage the system-level shared-resource contention for co-located applications by dynamically partitioning both the compute and memory resources without incurring high overhead. 
In particular, \sys exploits the regularity of DNN operators and hardware, where execution latency is highly correlated with the number of in-flight memory requests.

Specifically, \sys consists of 1) a lightweight hardware memory access monitoring and regulation engine, 2) an intelligent runtime system that manages the memory usage of each of the co-located applications dynamically, and 3) a priority-aware scheduler that selects the workloads to execute concurrently based on its user-assigned priority, latency target, and memory resource usage.
Our evaluation shows that \sys can improve overall QoS by up to 3.9$\times$ (1.8$\times$ on average), together with an improvement in system throughput of 2.3$\times$ and fairness of 1.3$\times$, compared to prior work that partitions compute resources only~\cite{Planaria}.
In summary, this paper makes the following contributions:
\begin{enumerate}[topsep=1pt,itemsep=-0.2ex,partopsep=1ex,parsep=1ex]
\item  We develop \sys, an adaptive contention management mechanism through co-design of hardware, runtime system, and scheduler to adaptively adjust the contentiousness of co-located DNN applications.
\item We implement the \sys hardware in RTL that dynamically monitors and regulates the memory access rates of DNN accelerators.
\item We design the \sys runtime system that adaptively configures \sys hardware based on the target latency and priority of the co-located workloads.
\item We build the \sys scheduler, a priority- and memory-contention-aware workload scheduler to select different DNN layers to run concurrently.
\item  We demonstrate the effectiveness of \sys in FPGA-based full system simulation and synthesize and place-and-route it using an advanced 12nm process technology.
Our results show that, compared to prior work, \sys significantly improves the performance of multi-tenancy execution over a range of deployment scenarios.

\end{enumerate}

}

\section{Background and Motivation}
This section discusses the need for multi-tenancy execution in today's DNN workloads, the challenges associated with multi-tenancy, along with prior work in this space.

\subsection{Modern DNN Applications}
With the growing popularity of deep learning, many modern applications employ a diverse set of DNN algorithms to support different functionalities.
For example, robotics~\cite{mavbench-micro2018,dl-robotics,lb-wayptnav} and AR/VR~\cite{illixr,edge-facebook,hda-hpca2021} applications need to detect nearby agents, track their movements, and predict the paths they will take, many of which are DNN-based and need to be performed concurrently.
Furthermore, these workloads differ greatly in their latency and bandwidth requirements, ranging from real-time tasks with strict QoS deadlines, such as eye tracking on AR / VR devices~\cite{edge-facebook}, to offline workloads that run only when the SoC is idle, e.g., identifying people and objects in a photo album~\cite{mlperf-inference-isca2020}.
In this case, multi-tenancy execution, where multiple workloads are executed simultaneously, is a natural solution to avoid overprovisioning of hardware resources while efficiently supporting concurrent execution.

\subsection{Challenges with Multi-Tenancy Execution}

While multi-tenancy execution generally improves hardware utilization, it also often faces performance degradation due to resource contention.
Even in the context of domain-specific accelerators, while each individual accelerator is generally designed in isolation, once integrated into an SoC, they typically share a number of system resources, including last-level cache (LLC), system bus, DRAM, and other I/O devices.
As a result, co-located applications running on accelerators can also compete for these shared resources, leading to significant performance degradation.

\begin{figure}[t]
    \centering
    \begin{subfigure}{.48\linewidth}
      \centering
      \includegraphics[width=\linewidth]{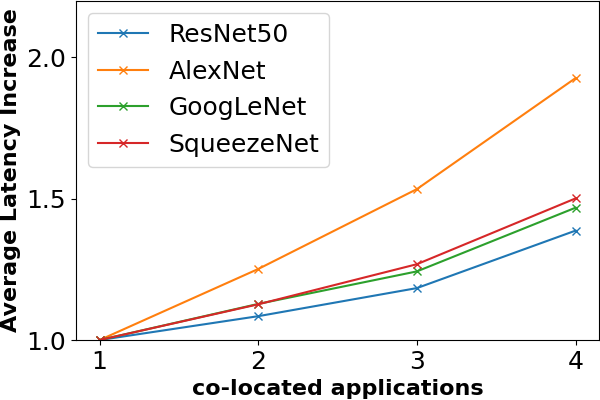}
      \caption{Average latency increase.}
       \label{fig:motiv_average_scale}
    \end{subfigure}
    ~
    \begin{subfigure}{.48\linewidth}
      \centering
      \includegraphics[width=\linewidth]{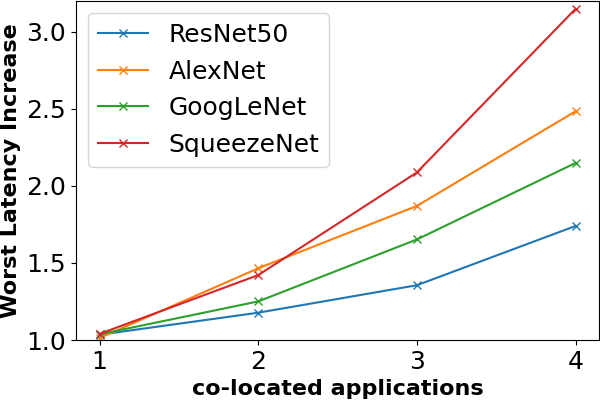}
      \caption{Worst-case latency increase. }
      \label{fig:motiv_worst_scale}
    \end{subfigure}
    \caption{Average and worst-case latency increase due to colocated DNN applications. When x=1, the workload is running in isolation with no performance degradation. When x=3, the workload is co-located with two other randomly selected workloads. We measure the end-to-end execution latency of 300 simulations with different starting time to calculate the average and worst latency across all runs.}
    \label{fig:motiv_scale}
    
\end{figure}

To understand the performance implications of multi-tenant DNN execution, we co-locate four different DNNs~\cite{resnet, alexnet, squeezenet, googlenet} on a state-of-the-art SoC with a DNN accelerator, general-purpose cores, shared system memory, and DRAM. These configurations are similar to commercially available SoCs such as NVIDIA's Xavier~\cite{nvdla-hotchips}.
We randomly dispatched them at different times to capture different patterns of interference between them. We measure end-to-end execution latency using our FPGA-based RTL evaluation.
Figure~\ref{fig:motiv_scale} shows the average and worst-case latency increases of multi-tenant execution.
Latency is normalized to the latency of the workload running in isolation where there is no contention.

We observe at least 40$\mathit{\%}$ latency increase across all workloads when they are co-located with the other DNNs.
In particular, AlexNet shows an almost 2$\times$ average latency increase when four applications are co-located, shown in Figure~\ref{fig:motiv_average_scale}, as AlexNet is quite sensitive to memory capacity as its latency is dominated by the memory-intensive fully connected layers.
We also observe significant performance variations across runs.
For example, the worst-case latency for SqueezeNet is more than 3$\times$ of its isolated execution, shown in Figure~\ref{fig:motiv_worst_scale}.
This is because the runtime of SqueezeNet is relatively short, making its latency quite sensitive to the behaviors of co-located workloads.
The worst case happens when the entire execution of SqueezeNet is co-located with memory-intensive layers.
Hence, while it is straightforward to co-locate multiple DNN workloads on the same hardware, the most important challenge is to meet the performance requirements across different co-running scenarios.

\subsection{Architectural Support for Multi-Tenancy}
\label{prior-work}

Table~\ref{tab:taxonomy} summarizes hardware and software techniques developed to support multi-tenancy execution in general-purpose processors and, more recently, domain-specific accelerators.
Multi-tenancy execution in traditional general-purpose cores is a well-studied topic in both software and hardware, dating back to the introduction of chip-multiprocessors. 
Generally, two lines of work have been proposed.

The first category focuses on \textit{static} mechanisms, where software or hardware is configured statically to adjust the memory access rates of contentious applications.
For example, QoSCompile~\cite{QoS-compile} statically throttles the contentious code regions, Locality-aware complication~\cite{CMP_cache} re-compiles applications to reduce multicore resource contention,
and Prophet~\cite{prophet} leverages accurate QoS prediction to identify safe co-location opportunities.
While these static approaches are lightweight and do not require runtime information, they are typically overly conservative for low-priority applications, as their memory requests are delayed regardless of whether the co-running applications are indeed sensitive to contention.

\begin{table}[t]
\centering
\scalebox{0.75} {
\begin{tabular}{ l | c | c || c | c | c}
\hline
                  & \multicolumn{2}{c||}{\textbf{General Purpose Hardware}}                                      & \multicolumn{3}{c}{\textbf{Domain Specific Hardware}}                                                                                                                                                                                                                      \\ \hline
\multirow{2}{*}{} & \multicolumn{1}{c|}{\multirow{2}{*}{\textbf{Software}}} & \multirow{2}{*}{\textbf{Hardware}} & \multicolumn{1}{c|}{\multirow{2}{*}{\textbf{\begin{tabular}[c]{@{}c@{}}Temporal\\ Partition\end{tabular}}}} & \multicolumn{2}{c}{\textbf{\begin{tabular}[c]{@{}c@{}}Spatial Partition\end{tabular}}}                                                                      \\ \cline{5-6} 
                  & \multicolumn{1}{c|}{}                                   &                                    & \multicolumn{1}{c|}{}                                                                                       & \multicolumn{1}{c|}{\textbf{\begin{tabular}[c]{@{}c@{}}Compute\\ Resource\end{tabular}}} & \textbf{\begin{tabular}[c]{@{}c@{}}Memory\\ Resource\end{tabular}} \\ \hline

 \hline
\textbf{Static} &
\makecell{Prophet~\cite{prophet},\\ QoSCompile\\~\cite{QoS-compile}, \\Locality-aware\\ Compilation\\\cite{CMP_cache}}
& \makecell{SplitL2~\cite{lastlinedefense},\\
PAR-BS~\cite{batch_scheduling},\\
FairQueuing\\~\cite{multicore_drambw},\\
STFM~\cite{STFM},\\
CoQoS~\cite{coqos}} &
\makecell{LayerWeaver\\~\cite{layerweaver}} & \makecell{HDA~\cite{hda-hpca2021}} & 
\makecell{CachePartition\\~\cite{cache_partitioning}\\{MAGMA~\cite{magma}}} \\
\hline
\makecell{\textbf{Dynamic}}& 
\makecell{
ReQoS~\cite{ReQoS},\\
CAER~\cite{CAER},\\ 
Kelp~\cite{kelp-hpca2019},\\
Baymax~\cite{Baymax},\\
Quasar~\cite{quasar-asplos2014},\\
Paragon~\cite{paragon-asplos2013}, \\
Thread-throttle\\~\cite{Thread_throttling_micro2010}
} &
\makecell{
FST~\cite{Ebrahimi2012FairnessVS},\\
QoS in CMP\\~\cite{QoSframework},\\
Utility-based\\
Partition~\cite{utilities_cache_partition},\\
Heracles\cite{Heracles} \\
FCSP~\cite{Fair_cache_share_pact04}}&
\makecell{Prema~\cite{PREMA}\\AI-MT~\cite{AI-MT}} &
\makecell{Planaria~\cite{Planaria}\\Veltair~\cite{Veltair}} & \textbf{\sys} \\
\hline
\end{tabular}
}
\caption{Taxonomy of multi-tenancy support in general-purpose hardware and domain-specific accelerators.}
\label{tab:taxonomy}
\end{table}


The second category targets \textit{dynamic} mechanisms, where runtime information is collected to decide when contention management mechanisms should be triggered and how aggressive these mechanisms need to be, either in software or hardware.
Most of the proposed dynamic approaches target general-purpose processors.
While these dynamic approaches are generally more involved with additional software/hardware modules to track runtime information, they can adaptively adjust the contentious nature of an application based on the amount of memory traffic that is actually occurring in the system, leading to improved QoS and throughput.

\begin{figure}[b]
    \centering
    \includegraphics[width=\linewidth]{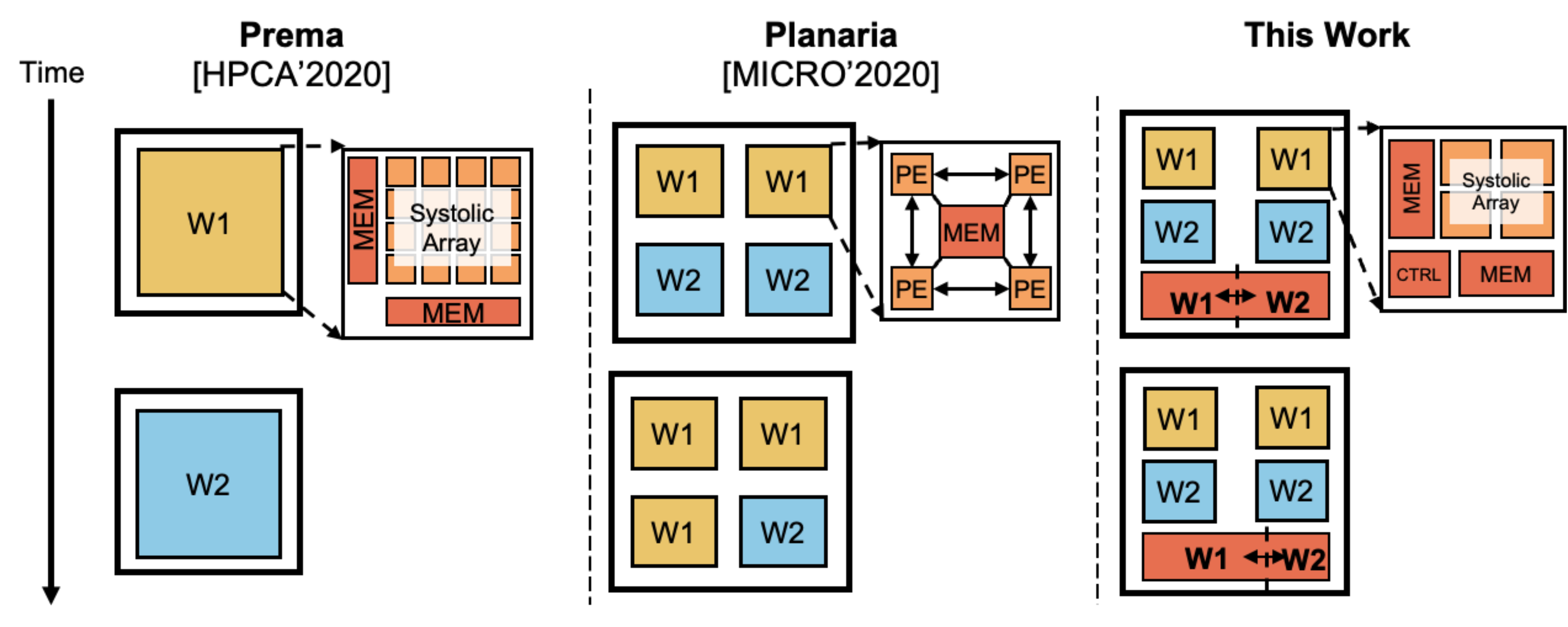}
    \caption{Comparison of the recently proposed multi-tenancy mechanisms. Prema~\cite{PREMA} supports temporal multi-plexing across different DNN workloads (W). Planaria~\cite{Planaria} focuses on dynamic partition of the compute resources at the pod granularity, where each pod consists of a fixed number of PEs and scratchpads. Different from prior works, \sys adaptively partitions \textit{both the compute and the shared memory resources} to satisfy the resource requirements of different workloads. }
    \label{fig:comparison}
\end{figure}

When it comes to domain-specific accelerators, Table~\ref{tab:taxonomy} further categorizes how shared resources are partitioned over time, i.e., temporal partitioning, or over space, i.e., spatial partitioning.
In spatial partitioning, we also separate whether compute resources, e.g., cores or processing elements in spatial arrays, or memory resources, e.g., shared cache and DRAM bandwidth, are partitioned spatially.
Previous work such as LayerWeaver~\cite{layerweaver}, Prema~\cite{PREMA}, and AI-MT~\cite{AI-MT} propose time-multiplexing DNN execution, statically or dynamically, although they suffer from low hardware utilization, especially when layers cannot utilize all available resources.
To improve hardware utilization, static partition techniques such as HDA~\cite{hda-hpca2021} and MAGMA~\cite{magma} propose spatial partitioning of compute or memory resources, although they cannot adapt to different dynamic scenarios due to their static nature.

Recently, dynamic spatial partition mechanisms have been proposed to further improve the co-location efficiency.
Figure~\ref{fig:comparison} highlights the key differences among recent efforts in this space.
In particular, Planaria~\cite{Planaria} and Veltair~\cite{Veltair} propose dynamic allocation of compute resources for co-running workloads.
Although they can adaptively allocate compute resources, e.g., processing elements for Planaria or CPU cores for Veltair, they are not able to dynamically manage memory resources, leading to low resource utilization and high thread migration overhead during compute resource repartition~\cite{Veltair}.
\sys overcomes these challenges by emphasizing the importance of memory-centric resource management by dynamically partitioning both compute and memory resources during execution.
\textit{To the best of our knowledge, \sys is the first work that adaptively partitions the memory resources to support multi-tenant execution in DNN accelerators.}

\section{\sys System}\label{architecture}

\begin{figure}[t]
    \centering
    \includegraphics[width=0.9\linewidth]{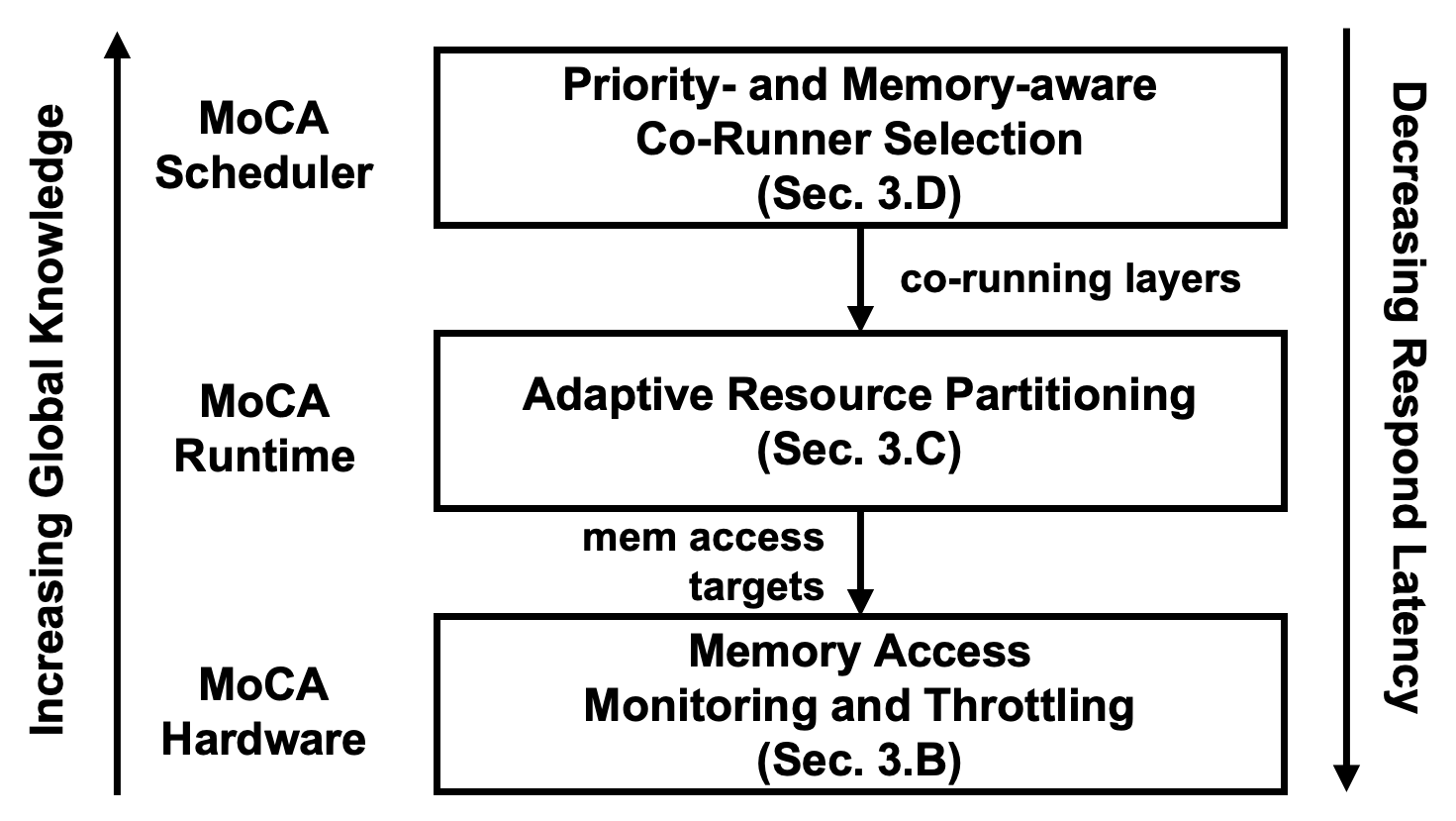}
    \caption{Overview of the \sys System.}
    \label{fig:moca_overview}
\end{figure}

We propose \sys, a memory-centric, adaptive DNN accelerator architecture that dynamically partitions both compute and memory resources to colocated DNN workloads to improve the multi-tenant performance of DNN accelerators. 
In particular, \sys dynamically manages the memory access rates of co-running workloads through hardware, runtime, and scheduler co-design to meet the QoS requirements of applications.
This section first provides an overview of the \sys system, followed by detailed discussions of \sys's hardware, runtime system, and scheduler components.

\subsection{\sys Overview}
\sys is a full stack system composed of 1) a lightweight hardware monitoring and throttling engine to control the memory accesses of the colocated DNNs, 2) an intelligent runtime that dynamically modulates the processing rate of each workload, and 3) a priority- and memory-aware job scheduler that selects the co-running workloads, as shown in Figure~\ref{fig:moca_overview}.
Starting at the bottom of the stack, the \sys hardware monitors the memory access counts of each individual accelerator tile and throttles execution if the hardware exceeds the allocated memory access targets over a certain amount of time.
The memory access target is set by the \sys runtime system, which detects interference across co-running jobs and adaptively repartitions the available resources if needed.
Moving up the stack, the co-running jobs are selected by the \sys scheduler, which takes the user-defined priorities and memory requirements of the network layers as input and chooses the co-running layers that meet the QoS requirements of the workloads.

Together, the hardware, runtime, and scheduler components of \sys reduce system-level contention  and deliver improved QoS for multi-tenant execution.
Next, we will discuss each component of the \sys system in more detail.

\subsection{\sys Hardware}
The responsibility of the \sys hardware is to dynamically monitor the memory access rate of each accelerator tile and throttle its execution if the target access counts have been reached.
Although dynamic memory throttling has been implemented in general-purpose systems before, to the best of our knowledge, \sys is the first work that demonstrates adaptive memory regulation in multi-tenant accelerators.
Unlike early multi-tenant accelerators that focus on compute resource partition~\cite{PREMA, Planaria, Veltair}, the \sys hardware builds on top of the decoupled access/execute nature of state-of-the-art DNN accelerators~\cite{tpu-isca2016} and controls accelerator's memory accesses independently without changing accelerator's compute engine. 

\begin{figure}[t]
    \centering
    \includegraphics[width=0.8\linewidth]{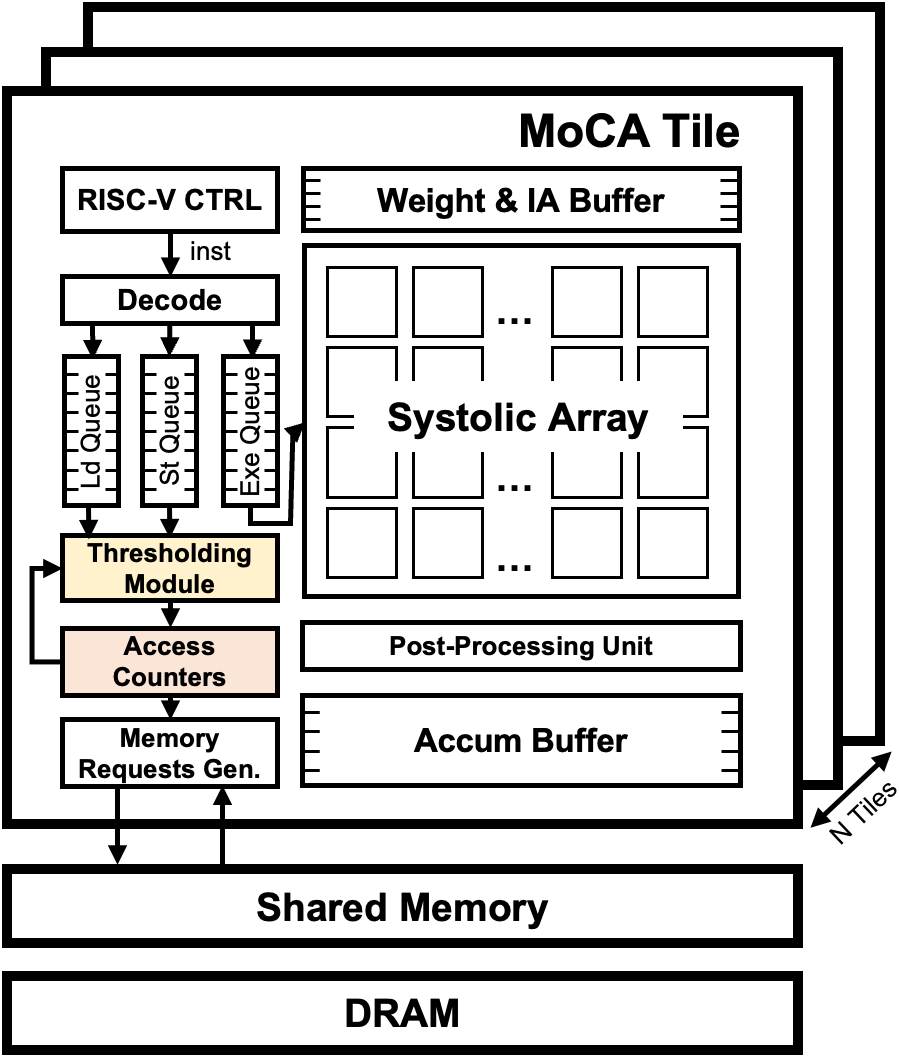}
    \caption{The \sys DNN accelerator that supports dynamic memory access monitoring and throttling. The \hw monitors the memory accesses that have been issued from each of the accelerator tiles and throttle their memory accesses if it exceeds the limit.}
    \label{fig:arch_overview}
\end{figure}

Figure~\ref{fig:arch_overview} shows the hardware microarchitecture of the proposed \sys DNN accelerator.
In addition to the standard systolic array and buffers (to store weights (W), input activations (IA), and output activations (OA) in DNN), \sys adds two new hardware modules between the ld/st queues and the memory request generation engine to monitor and throttle its own locally-generated memory accesses to the shared memory resource.
Specifically, \sys implements an \textit{Access Counter to locally track its memory access counts during the monitored time \textit{window}, and a \textit{Thresholding Module} to prevent accelerator from further generating memory requests if the \textit{Access Counter} value exceeds its \textit{threshold load} during that time \textit{window}, which is configured by the MoCA runtime system.}

We implement our hardware based on the Gemmini infrastructure~\cite{gemmini-dac}, an open-source RISC-V-based DNN accelerator infrastructure that supports TPU-like decoupled access/execute execution.
The systolic array and buffers are generated directly from Gemmini, and we implement the \sys-specific access monitoring and throttle engines in Chisel RTL~\cite{chisel2012-dac}.
Specifically, the \textit{Access Counter} monitors accelerator memory requests to check whether the current execution is overly utilizing memory than its target number of memory requests configured by MoCA runtime system.

Based on the rate at which the accelerator issues load requests, the thresholding module determines how many cycles ``bubbles'' should be inserted to block further memory requests.
Bubbles prevent the accelerator from making additional load requests, allowing \sys to reduce memory contention while providing each accelerator with the number of memory accesses allocated by the MoCA runtime.
Whenever the access counter raises an alert that its accumulated counter exceeds threshold, the \sys hardware begins to stall, until its status is updated by the \sys runtime.
Unlike previous work that required a significant redesign of DNN accelerators to support flexible resource partitioning~\cite{Planaria}, the \sys hardware engines can be implemented as lightweight finite-state machines and counters without incurring significant overhead.

\subsection{\sys Runtime}\label{runtime section}
The \sys runtime is responsible for dynamically detecting system-level interference and setting limits on the memory access rates of accelerators to resolve contention if necessary.
Existing multi-tenancy solutions focus on compute resource partitioning. While easy to manage, these solutions tend to incur non-negligible repartitioning latency due to high thread migration overhead, as observed in prior work~\cite{Planaria, Veltair}.
Instead, \sys's full-stack design exposes the hardware states to \sys's runtime so that it can dynamically partition both memory and compute resources based on the demand of applications and the cost of re-partitioning.
In fact, \sys's runtime triggers the compute resource partition much less frequently to avoid its high overhead.
\begin{algorithm}[t]
\caption{Latency Estimation in \sys Runtime} \label{alg:runtime_prediction}
\begin{algorithmic}[1]
    \LeftComment{\% Estimate latency for COMPUTE layers \% }
    \If{$Layer_{i}$ is COMPUTE}
        \State $Total\_MAC_{i} \leftarrow$ calc\_MAC\_count($Layer_{i}$)  
        \LeftComment{\% Calculate compute-only time \% }
        \LeftComment{\% Ideal compute time with 100\% PE utilization \%}
        \State $Compute\_ideal_{i} \leftarrow Total\_MAC_{i} / num\_PEs$
        \\
        \LeftComment{\% Calculate memory-only time \% }
        \LeftComment{\% Total traffic to shared L2 \%}
        \State $Total\_MEM_{i} \leftarrow Total\_load_{i}+Total\_store_{i}$
        \LeftComment{\% Subset of traffic to DRAM \%}
        \State $From\_DRAM_{i}  \leftarrow (Weight +  Output + Bias)\_size_{i}$
        
        \LeftComment{\% If input can be cached, no need to reload from DRAM. Otherwise, reload. \% }
        \If{$Image\_size_{i} > Cache\_size$}
            \LeftCommentFirst{0.87cm}{\% If input activation got evicted \%}
            \State $From\_DRAM_{i}\ \textrm{+=}\ Image\_size_{i}$
        \EndIf 
        \If{$Per\_tile\_size_{i} > Cache\_size$}
            \LeftCommentFirst{0.87cm}{\% If tile size exceeds cache size, it got evicted \%}
            \State $From\_DRAM_{i} \textrm{+=} Tiling\_factor \cdot Tile\_size$
        \EndIf 
        \LeftComment{\% Consider both L2 and DRAM transaction time \%}
        \State $Memory\_ideal_{i} \leftarrow \frac{From\_DRAM_{i}}{DRAM\_BW} + \frac{Total\_MEM_{i}}{L2\_BW}$
        \\
        \LeftComment{\% Estimate overall latency from compute \& memory time, considering compute-to-memory ratio \% }
        \State $Prediction_{i}\!\! \leftarrow\!\! Max(\!Compute\_ideal_{i}, Memory\_ideal_{i}\!)$ 
        \State $ + Min(Compute\_ideal_{i}, Memory\_ideal_{i})\! \cdot\! overlap\_f$ %
        
    \EndIf
    \\
    \LeftComment{\% \textrm{Estimate latency for MEM layers} \% }
    \If{$Layer_{i}$ is MEM}
    \State $\small Total\_Mem_{i}\! \leftarrow\!  InputAB\_size_{i}\! +\! Output\_size_{i}$
    \State $\small From\_DRAM_{i}\leftarrow InputB\_size_{i} + Output\_size_{i}$
    \State $Prediction_{i} \leftarrow \frac{From\_DRAM_{i}}{DRAM\_BW} + \frac{Total\_MEM_{i}}{L2\_BW}$
    \EndIf
\end{algorithmic}
\end{algorithm}

\begin{algorithm}[t]
\caption{\sys Contention Detection and HW Update} \label{alg:mem_rate_compute}
\begin{algorithmic}[1]
\LeftComment{\% Get latency and memory usage of the remaining layers using Algorithm~\ref{alg:runtime_prediction} \% }
    \State $remain\_prediction \leftarrow Alg1(Layers)$
    \For{$Layer_{i}$ in Layers}
        \State $(From\_DRAM_{i}\!, \!Total\_MEM_{i}\!, \!Prediction_{i}\!) \!\!\! \leftarrow \!\!\! Alg1\!(Layer_{i}\!)$
        \State $BW\_rate_{i} \leftarrow From\_DRAM_{i} / Prediction_{i}$
        \State$slack \leftarrow time\_left\_to\_target$
        \LeftComment{\% Dynamic priority score update \%}
        \State $priori\_score \!\leftarrow\! user\_priority_{i} \!+\! \frac{remain\_prediction}{slack}$
        \LeftComment{\% Latency prediction of remaining layers of NN \%}
        \State $remain\_prediction\ \textrm{-=}\ Prediction_{i}$
        \\
        \LeftComment{\% Look up co-running applications' memory usage in scoreboard \% }
        \For{$App_{j}$ in $other\_Running\_Apps$} 
                \LeftCommentFirst{0.87cm}{\% Add current memory BW usage per workload\%}
                \State $other\_BW\_rate\ \textrm{+=}\ MEM\_BW(App_j)$
                \LeftComment{\% Weighted sum using dynamic priority score \%}
                \State $\small weight\_sum\ \textrm{+=}\ score(App_j)\! \cdot\! MEM\_BW(App_j)$ 
        \EndFor
        \\
        \LeftComment{\% Whether the current system's total memory demand is bigger than the maximum BW \% }
        \State $overflow\!\leftarrow\! BW\_rate_{i} \!\!+\!\!other\_BW\_rate \!-\! DRAM\_BW\_MAX$
        \LeftComment{\% Contention detected (real-time detection) \% }
        \If{$overflow > 0$} 
            \State $curr\_weight\_sum\! \leftarrow\! priori\_score \! \cdot\! BW\_rate_{i}$
            \LeftComment{\% Allocate memory w/ dynamic priority score \%}
            \State $BW\_rate_{i}\ \textrm{-=}\ \frac{overflow * weight\_sum}{curr\_weight\_sum + weight\_sum}$
            \LeftComment{\% Update prediction based on allocation \%}
            \State $Prediction_{i} \leftarrow BW\_rate_{i}  \cdot From\_DRAM_{i}$
            
            \\
            \LeftComment{\% Set HW config (dynamic memory partition) \%}
            \State $threshold\_load_{i}\! \leftarrow\! Total\_MEM_{i} / Num\_tile_{i}$
            \State $window_{i} \leftarrow Prediction_{i} / Num\_tile_{i}$
            
        \Else 
            \LeftCommentFirst{0.87cm}{\% Contention not detected, no throttling \%}
            \State $threshold\_load_{i} \leftarrow 0$,\quad  $window_{i} \leftarrow 0$
        \EndIf
        \State UpdateScoreboard($this\_App, BW\_rate_{i}$)
        \State ConfigureHW($window_{i}$, $threshold\_load_{i}$)
        \State runLayer($Layer_{i}$)
    \EndFor
\end{algorithmic}

\end{algorithm}

Specifically, \sys's runtime consists of two components.
The first part is the performance prediction of co-running layers based on available hardware resources, and the second part is the contention detection module and the hardware re-partition module which activates when contention occurs.
To accurately estimate the performance and memory requirements of the co-running layers, the \sys runtime captures the data movement costs across the full memory system.
If the memory bandwidth requirement of the co-running layers is greater than the available bandwidth in the system, \sys's runtime declares that contention is detected.
In that case, \sys's runtime reconfigures the \sys hardware to either limit the memory access rate or, in rare cases, repartition the compute engines, based on both user-defined priorities and target latencies of co-running applications.



Algorithm~\ref{alg:runtime_prediction} illustrates the latency estimation algorithm in the \sys runtime.
Unlike compute-oriented latency estimation in prior multi-tenant solutions~\cite{PREMA, Planaria, Veltair}, \sys considers the movement of data across the full memory system, including both shared memory (i.e., L2) and DRAM, allowing fast and accurate performance prediction based on different memory resources. 
Specifically, \sys categorizes DNN layers into two types: \compute layers that have high arithmetic intensity (such as convolutional layers or fully-connected layers), and \mem layers that exhibit little data reuse and are composed of memory-bandwidth-bound operators (such as residual additions and max-poolings which cannot be fused with CONV layers).
\texttt{overlap\_f} is a parameter that represents system's ability to overlap compute and memory operations, which can be varied ($0<overlap\_f<1$) by accelerator design. We provide a tuning utility that determines the optimal value of $f$ for an SoC using data collected by running a few DNN layers before starting inference queries.
For each layer of DNN networks, Algorithm~\ref{alg:runtime_prediction} calculates the total number of multiply and accumulate operations (MAC) to compute based on the dimensions of the layer, together with the cost of data movement between different levels of the memory hierarchy, such as DRAM and L2.
Our validation shows prediction error within 10\% of measured runtimes across networks and layers, consistent with other analytical models for DNN hardware such as Timeloop~\cite{timeloop2019-ispass}.

Algorithm~\ref{alg:mem_rate_compute} shows how the \sys runtime sets parameters for the \sys hardware engine. 
We first use Algorithm~\ref{alg:runtime_prediction} to calculate the memory requirement and estimate the latency of the current layer without interference, which are deterministic based on the layer and hardware configuration.
This value is used to calculate the workload's score. The score represents the relative importance of the workload, and is computed using both the workload's priority and whether enough time remains to finish the entire network within its target deadline. 
The latency prediction (\texttt{remain\_prediction}) is updated for every layer, since the scores are updated based on the layers that have not been executed.
The MoCA hardware engine is reconfigured each time the dynamic scores are updated, since relative importance and system contention would change over time due to changes in co-running workload.
Dynamic priority in \sys considers both the SLA target and user-defined static priority. In Algorithm~\ref{alg:mem_rate_compute}, \texttt{priori\_score} is the dynamic score used in \sys, while \texttt{slack} calculates the time remaining to hit the SLA target, and \texttt{user\_priority} is the static priority.
Thus, when memory contention is detected and memory resources get partitioned, \sys prioritizes applications with higher priorities and applications with less time left to meet their targets.

\subsection{\sys scheduler} \label{scheduler}

Moving up the stack, we also need a workload scheduler to intelligently select co-running applications that can be executed concurrently in the system.
In particular, the multi-tenant scheduler takes the following requirements into consideration:
\begin{enumerate}
\item  The scheduler needs to be aware of the existence of different user-given priorities and has to react accordingly. 
\item  The scheduler needs be aware of both the memory and compute resource requirements of workloads.
\item  The scheduler algorithm needs to be light-weight and simple enough to be computed in real-time without incurring significant overhead.
\end{enumerate}
Existing multi-tenant schedulers lack the consideration of memory resources~\cite{PREMA, Planaria} as well as user-defined priority and latency targets~\cite{hda-hpca2021, magma}, making them unsuitable for handle a diverse set of workloads under different deployment scenarios. 


\begin{algorithm}[!t]
\caption{\sys Scheduler for Multi-tenant Execution} \label{alg:scheduler_algorithm}
\begin{algorithmic}[1]
\LeftComment{\% During the scheduling period, calculate and update the score based on the user-given priority \%}
\State $curr = current\_time()$
    \For{each \textit{$Task_{i}$} in \texttt{TaskQueue}}
        \State $Score_{i} \leftarrow user\_given\_priority_{i}$
        \State $WaitingTime_{i} \leftarrow curr - DispatchedTime_{i}$
        \State $Slowdown_{i} \leftarrow \frac{WaitingTime_{i}}{EstimatedTime(Task_{i})}$ 
        \State $Score_{i}\ \textrm{+=}\ Slowdown_{i}$
        \LeftComment{\% Flag the task if it is memory intensive \%}
        \If{$EstimatedAvg\_BW_{i} > 0.5\times DRAM\_BW$}
            \State $isMemIntensiveTask[i] \leftarrow $ True
        \Else
            \State $isMemIntensiveTask[i] \leftarrow $ False
        \EndIf
    \EndFor
    \\
\LeftComment{\% Populate ExQueue based on the updated score \%}
    \State $ExQueue \leftarrow$ [$Task_{i}$ if $Score_{i} > Threshold$]
    \State $ExQueue.sort()$
    \\
\LeftComment{\% Form co-running tasks based on their scores and memory intensiveness \%}
    \State {$groupTask = []$}
    \While {$resourceAvailable$}
        \State $currTask = ExQueue.pop()$
        \State $groupTask.append(currTask)$
        \If {$isMemIntensiveTask(currTask)$}
            \State $coTask = findNonMemIntensiveTask(ExQueue)$
            \State $groupTask.append(coTask)$
        \EndIf
    \EndWhile \\
    \Return $groupTask$
    
\end{algorithmic}
\end{algorithm}

To meet these requirements, we design the \sys scheduler, described in Algorithm~\ref{alg:scheduler_algorithm}.
In particular, the \sys scheduler considers both the user-defined priority of each task and how long the task has been dispatched when it selects co-running workloads.
Specifically, \sys uses a \texttt{TaskQueue} to store the dispatched workloads.
Each task can be defined using either model granularity or as groups of layers if there are significant differences in the compute-to-memory ratio between layers within the model.
Each of the task queue entries includes the Task ID, the dispatch time to \texttt{TaskQueue}, the current status of the workload, the user-given priority, and the target latency. 

During each round of scheduling, the \sys scheduler compares the \texttt{scores} of each task, which considers both the user-defined priority of each task and how long the task has been waiting in the queue, selects the highest ranked tasks and puts them in the execution queue.
In addition, the \sys scheduler also considers the memory resource requirements of different tasks during its scheduling.
In the case of memory-bound layers that require extensive memory usage, the \sys scheduler will co-schedule these layers with other layers later in the queue with lower memory requirements.

\subsection{Limitations of \sys}
\sys leverages the regularity of both DNN workloads and DNN hardware, where it assumes dense DNN workloads and does not exploit the sparsity of data. 
It also best targets spatial hardware architecture where the performance of executing dense DNN layers is largely determined by the available compute and memory resources.
This is because if sparsity is considered in hardware, it can be challenging to estimate the memory requirements of the DNN layers during runtime. 
However, \sys can be augmented with an accurate performance and memory resource predictor of sparse DNNs to support sparse DNN accelerators.

\section{Methodology}\label{methodology}
This section details \sys's implementation, our workloads and metrics used for our evaluation, as well as the baseline co-location solutions against which we compare the effectiveness of \sys.

\subsection{\sys implementation}

We implement \hw using the Chisel RTL language~\cite{chisel2012-dac} on top of the Gemmini~\cite{gemmini-dac} infrastructure, a systolic-array-based DNN accelerator without multi-tenancy support.
We evaluate performance on full, end-to-end runs of DNN workloads using FireSim, a cycle-accurate, FPGA-accelerated RTL simulator~\cite{firesim}.
We also synthesize our hardware implementations on the GlobalFoundries 12nm process to evaluate the area overhead.

Table~\ref{tab:hardware_config} shows the SoC configuration we use in our evaluations of \sys, similar to the configurations in modern SoCs~\cite{nvdla-hotchips}.
We configure Gemmini, TPU style systolic-array based accelerator generator, to produce homogeneous eight separate accelerator tiles on the same SoC, each of which can run a different DNN workload.
Multiple accelerator tiles can also cooperate to run different layers/regions of the same DNN workload.
Each accelerator tile is equipped with a 16x16 weight-stationary systolic array for matrix multiplications and convolutions, as well as a private scratchpad memory to store W, IA and OA.
All tiles also share access to the shared memory subsystem, including a shared last-level cache and DRAM, which is the main source of contention with co-located workloads.

The \sw and scheduler are implemented in C++ and run on top of a full Linux stack. \sys uses a lightweight software look-up table for the scoreboard that is used to manage the bandwidth usage of each application and queues to keep track of the tasks dispatched.
The algorithms used to calculate the hardware configuration on the \sw and the \sys scheduler are also implemented in software with little overhead observed.

\begin{table}[t]
\centering
\scalebox{1.2} {
\begin{tabular}{ l | r }
\hline
\textbf{\makecell{Parameter}} &    \textbf{\makecell{Value}} \\
\hline
\hline
Systolic array dimension (per tile) & 16x16 \\
\hline
Scratchpad size (per tile) & 128KiB \\
\hline
Accumulator size (per tile) & 64KiB \\
\hline
\# of accelerator tiles & 8 \\
\hline
Shared L2 size & 2MB \\
\hline
Shared L2 banks & 8 \\
\hline
DRAM bandwidth & 16GB/s \\
\hline
Frequency & 1GHz \\
\hline
\end{tabular}
}
\caption{SoC configurations used in the evaluation.}
\label{tab:hardware_config}
\end{table}

We synthesize and place-and-route \sys-enabled accelerator. We use Cadence Genus with GlobalFoundries 12nm process technology for synthesis. For place-and-route, we use Cadence Innovus.

\subsection{Workloads}
\paragraph{Benchmark DNNs} In our evaluations, we choose seven different state-of-the-art DNN inference models, including SqueezeNet~\cite{squeezenet}, GoogLeNet~\cite{googlenet}, AlexNet~\cite{alexnet}, YOLOv2~\cite{yolov2}, YOLO-lite~\cite{yololite}, Keyword Spotting~\cite{kws-res26}, and ResNet~\cite{resnet}. 
These DNN models represent a diverse set of DNN workloads, with different model sizes, DNN kernel types, computational and memory requirements, and compute-to-memory trade-offs.
We grouped workloads into workload sets based on the DNN model size to capture any different behaviors across different sets. The workload classification is based on ~\cite{Planaria}, which is one of our baselines.
Table~\ref{tab:workloads} shows the DNN benchmarks classified by size. Workload set-A is the group of lighter models, whereas Workload set-B groups heavier models. Workload set-C contains both, which is the mixture of set-A and set-B.

\paragraph{Multi-tenant workload sets}
To generate multi-tenant scenarios, we select N different inference tasks randomly dispatched to the system~\cite{mlperf-inference-isca2020}, where N ranges from 200 to 500.
We assign user-defined, static priority levels within the range 0 to 11, and follow the distribution on~\cite{google_data_center_trace, google_data_center_modeling}, consistent with the methodology used for Prema and Planaria. 
\sys considers the priority levels when it schedules and dynamically adjusts the different layers.

\paragraph{QoS targets}\label{qos target} We set our baseline QoS based on~\cite{edge_benchmark} since each of our accelerator tiles is close to an edge device.
In addition to the baseline QoS, we adjust our latency target to 1.2$\times$ and 0.8$\times$ QoS, which increases and decreases the latency target by 20\%, to evaluate how \sys reacts with different latency targets. QoS-H (hard) denotes 0.8$\times$ QoS, which is more challenging to achieve. QoS-L (light) is 1.2$\times$ QoS, which is lighter load. QoS-M stands for the baseline QoS target.

\begin{table}[t]
\centering
\scalebox{1.1} {
\begin{tabular}{l | l | l | l }
\hline
\textbf{\makecell{Workload}} & \textbf{\makecell{Model\\size}} & \textbf{\makecell{Domain}} & \textbf{\makecell{DNN models}} \\
\hline
    \multirow{3}{*}{\makecell{Workload\\set A}} 
    & \multirow{3}{*}{Light} 
    & Image Classification & \makecell{SqueezeNet~\cite{squeezenet}} \\\cline{3-4}
    & & Object Detection & \makecell{Yolo-LITE~\cite{yololite}} \\\cline{3-4}
    & & Speech Processing & \makecell{KWS~\cite{kws-res26}} \\\hline
    
    \multirow{2}{*}{\makecell{Workload\\set B}} 
    & \multirow{2}{*}{Heavy} 
    & Image Classification & \makecell{GoogleNet~\cite{googlenet},\\AlexNet~\cite{alexnet}, \\ResNet50~\cite{resnet}} \\\cline{3-4}
    & & Object Detection & \makecell{YoloV2~\cite{yolov2}} \\\hline
    \makecell{Workload\\set C} & Mixed & All & \makecell{All} \\
\hline
\end{tabular}
}
\caption{Benchmark DNNs used in evaluation and workload sets based on model size.}
\label{tab:workloads}
\end{table}

\subsection{Metrics}\label{subsec:metrics}
We quantify the effectiveness of \sys-enabled workload colocation using the metrics suggested in \cite{perf_metric_multiprogram}, including the percentage of workloads for which we satisfy the Service Level Agreement (SLA), the throughput of the colocated applications, and the fairness of the \sys's strategy to manage shared resources. 
The latency of each workload is measured from the time it is dispatched to the system till it finishes and commits, which includes both the time it waits in the task queue and its runtime.


\paragraph{SLA satisfaction rate}
We set the SLA target for each workload based on the three different QoS levels we defined above in Section~\ref{qos target} paragraph `QoS targets' 
(we use QoS and SLA targets interchangeably.)
In addition to the overall SLA satisfaction rate, we also measure the SLA satisfaction rate for each of the priority groups to highlight the effectiveness of \sys.

\paragraph{Fairness}
\textit{Fairness} metric measures the degree to which all programs have equal progress.
The fairness metric evaluates \sys's priority scoring method used in the \sys scheduler and dynamic partitioning memory bandwidth in the \sys runtime.
Here, we use $\mathit{C_i}$ to denote the cycle time of the i-th workload, $\mathit{single}$ represents only one workload running on the SoC, and $\mathit{MT}$ represents the multi-tenant execution.
Similar to other previous work on DNN multi-tenancy support~\cite{PREMA,Planaria}, we use a generalized version of \textit{fairness} that measures \textit{proportional progress} (PP) defined as follows:
\begin{equation} \label{priority_metric}
\mathit{PP}_\mathit{i} = \frac{\frac{\mathit{C_i^{single}}}{\mathit{C_i^{MT}}}}{\frac{\mathit{Priority_i}}{\sum_{\mathit{j=1}}^{\mathit{n}} \mathit{Priority_j}}}, \quad \quad \mathit{Fairness} = \mathit{min_{i,j}} \frac{\mathit{PP_i}}{\mathit{PP_j}}
\end{equation}

\paragraph{Throughput}
We also analyze the total \textit{system throughput} (STP) to evaluate the effectiveness of modulating the memory access rate of \sys in increasing the overall utilization of hardware resources. STP is defined as the system throughput of executing $\textit{n}$ programs, which is defined by summing each program's normalized progress, which ranges from 1 to $\textit{n}$.
Increasing STP requires maximizing overall progress when co-locating multiple applications. 
\begin{equation} \label{STP_metrics}
\mathit{STP} = \sum_{i=1}^{n} \frac{\mathit{C}_\mathit{i}^{\mathit{single}}}{\mathit{C}_\mathit{i}^{\mathit{MT}}}
\end{equation}

\subsection{Baseline}
To evaluate the effectiveness of \sys, we compare with three different baseline: 
\begin{enumerate}
    \item Prema~\cite{PREMA}, which supports multi-tenancy execution through a time-multiplexing DNN accelerator across different DNNs with a priority-aware scheduling algorithm;
    \item Static compute partitioning, which exploits spatial co-location of multiple workloads but does not repartition the resource during runtime;
    \item Planaria~\cite{Planaria}, which spatially co-locate multiple DNNs by dynamically partitioning the compute resources with a fixed compute vs. memory resource ratio;
\end{enumerate}
In addition, instead of the layerwise granularity to reconfigure resources, we break down DNN networks into layer blocks, which consists of multiple layers, and reconfigure at the layer-block granularity, as recent work demonstrate layer-block granularity delivers supreme performance~\cite{Veltair}.
For fair comparison, we compare \sys with the stated baselines on the same hardware configuration.

\begin{figure}[t]
    \centering
    \includegraphics[width=\linewidth]{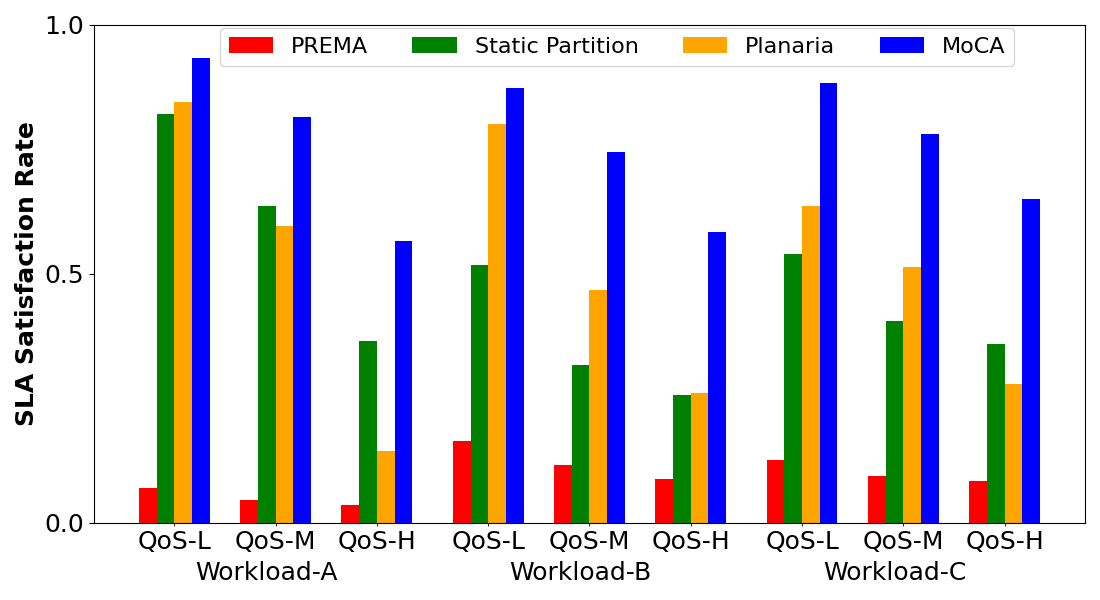}
    \caption{\sys's SLA satisfaction rate improvement over multi-tenancy baselines with different QoS targets (QoS-L: light latency target, QoS-M: medium latency target, QoS-H: hard latency target) and different DNN workload sizes (Workload-A: light models, Workload-B: heavy models, Workload-C: all models).}
    \label{fig:sla_overall}
\end{figure}

    
    

\section{Evaluation}\label{result}
{\rm
In this section, we evaluate the effectiveness of \sys-enabled multi-tenant execution in comparison to the three baselines discussed in Section~\ref{methodology}.
In particular, we compare the effectiveness of \sys to Prema~\cite{PREMA} and Planaria~\cite{Planaria}, both of which are recent works improving DNN multi-tenant execution.

We demonstrate that \sys meets the target QoS of different workloads with different priority levels, and we compare our fairness metrics to other baseline solutions.
In addition to increasing SLA satisfaction rates, \sys also improves resource utilization and system throughput, across a wide variety of workload scenarios with different DNN models and QoS requirements.
Finally, in addition to the above performance improvements, we also demonstrate that implementing the \sys's hardware components imposes only a small area overhead on DNN accelerators.



\begin{figure*}[t]
    \centering
    \begin{subfigure}{.32\textwidth}
      \centering
      \includegraphics[width=\linewidth]{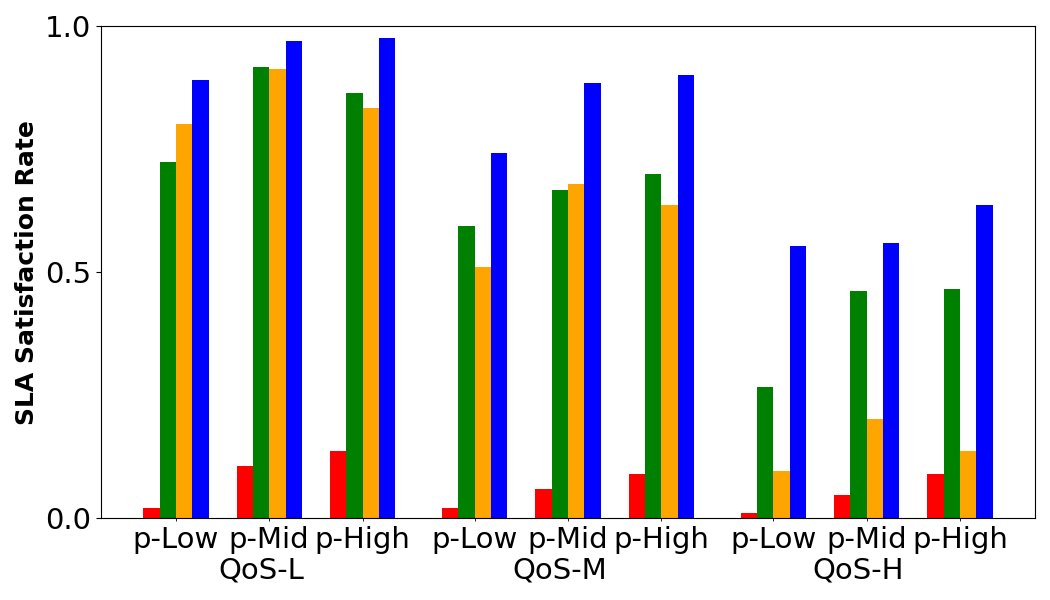}
      \caption{Workload-A}
       \label{fig:sla_light}
    \end{subfigure}
    ~
    \begin{subfigure}{.32\textwidth}
      \centering
      \includegraphics[width=\linewidth]{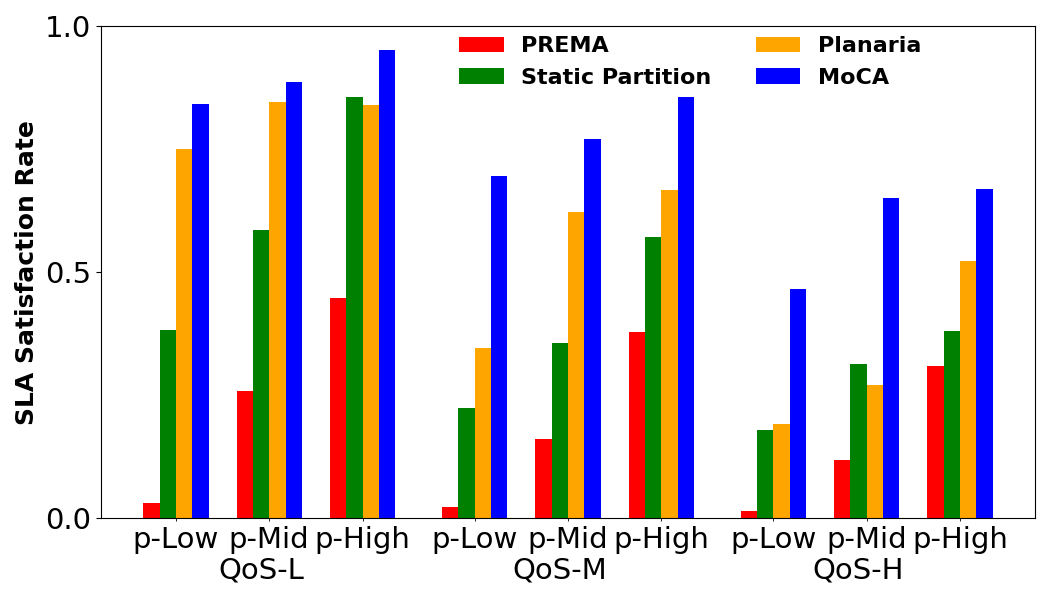}
      \caption{Workload-B}
      \label{fig:sla_heavy}
    \end{subfigure}
    \begin{subfigure}{.32\textwidth}
      \centering
      \includegraphics[width=\linewidth]{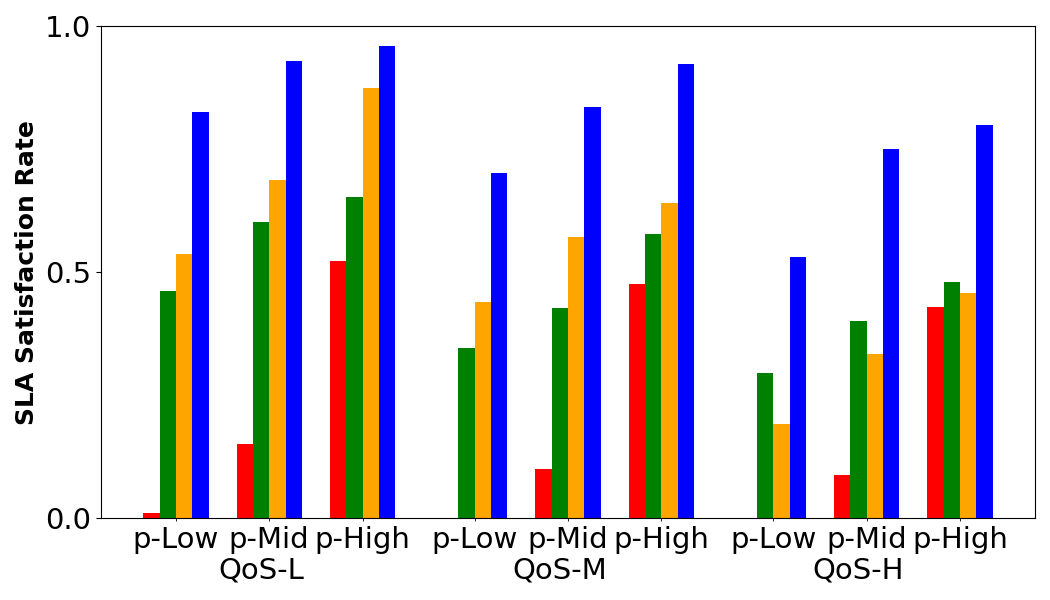}
      \caption{Workload-C}
      \label{fig:sla_mix}
    \end{subfigure}
     \caption{SLA satisfaction rate breakdown of different priority groups (p-Low: low priority, p-Mid: mid priority, p-High: high priority) and different DNN workload sizes (Workload-A: light models, Workload-B: heavy models, Workload-C: all models).}
    \label{fig:sla_priority}
    
\end{figure*}

\subsection{SLA satisfaction rate}
We evaluate three different set of workloads (listed in Table~\ref{tab:workloads}) with three different QoS targets (Hard: QoS-H, Medium: QoS-M, and Light: QoS-L), resulting in nine different runtime scenarios in  total, and measure the SLA satisfaction rate for each of them in comparison to three baselines.
As shown in Figure~\ref{fig:sla_overall}, \sys consistently outperforms all our baseline multi-tenancy execution mechanisms, for all the scenarios tested.
Across all scenarios, compared to Prema, \sys achieves an 8.7$\times$ geometric mean improvement, up to a maximum of 18.1$\times$.
\sys's SLA satisfaction rate shows a 1.8$\times$ geometric mean improvement than the static partitioning baseline, and up to 2.4$\times$ higher in the best cases.
Finally, \sys achieves up to a 3.9$\times$ higher SLA satisfaction rate than Planaria, and a 1.8$\times$ higher geometric mean.

\sys's improvement over the baseline solutions is most pronounced for the QoS-H scenario, where the SLA target is the hardest to achieve, demonstrating the effectiveness of \sys in adaptively partitioning resources to meet the performance targets in multi-tenant scenarios.
For more lenient targets, Planaria performs better than the static baseline, but its performance shows a steep degradation as the required QoS level increased. 
This is especially true for smaller DNN models, such as light models in Workload-A, where the number of resource reconfigurations increases, causing the cost of thread migrations in compute core repartition to be significantly more pronounced.
We observe that thread migration in compute thread repartitioning takes 1M cycles on average due to thread spawning and synchronization, similar to what was observed in recent work~\cite{Veltair}, which poses significant overhead to relatively short-running smaller models.
On the other hand, memory-partitioning in \sys only takes 5-10 cycles to reconfigure the DMA’s issue rate. Hence, \sys triggers memory repartitioning more frequently than compute repartitioning to avoid its high overhead.


On the other hand, when heavier workloads are grouped together, as in Workload-B, Planaria performed better than the static baseline, but its improvement is not as great as with \sys, especially as the QoS requirement tightens.
This is because the majority of Workload-B's workloads are memory intensive, undesirably degrading the performance of co-running layers.
Therefore, solutions that attempt to maintain the performance of these heavier workloads by reconfiguring the compute resources only at runtime can trigger frequent reconfigurations without much benefit, as slowdowns caused by memory contention cannot be resolved by reconfiguring only the compute resources.

Instead, \sys dynamically detects memory contention and modulates memory access, enabling it to adapt to the requirements of large colocated DNN models more effectively than compute resource reconfiguration.
In addition, reconfiguring \sys hardware does not require expensive thread migration, as it only requires issuing new hardware configuration commands to the accelerator.
Thus, \sys achieves better performance than prior works with its intelligent memory contention management and infrequent thread migration overhead.

\subsection{Priority analysis}
We further analyze SLA satisfaction results by dividing them into different priority groups based on user-given priority scores.
We classify workloads priority based on the distribution from \cite{google_data_center_trace}, where priority scores range from 0 to 11.
For visualization purposes, we grouped priorities into several categories: priority-low (p-Low) for scores 0 to 2, priority-mid (p-mid) for scores 3 to 8 and priority-high (p-high) for scores 9 to 11.
Figure~\ref{fig:sla_priority} shows the satisfaction rate grouped by different priority levels.
For the p-High groups, \sys improve the SLA satisfaction rate to 4.7$\times$ over Planaria (for Workload-A, QoS-H), 1.8$\times$ (for Workload-C, QoS-H) over the static partitioning baseline and 9.9$\times$ over Prema (for Workload-A, QoS-M).
\sys and the baselines all show a general trend to increase SLA satisfaction rates as workload priorities increased.
In particular, \sys is the only system that consistently delivers reliable performance across all workload, QoS, and priority scenarios, while other systems show performance degradation in difference cases.
For example, for Workload-A, Planaria achieves worse performance for p-High workloads than p-Mid workloads, as aggressive claiming more compute resources for high-priority workloads leads to expensive thread migration costs with little performance benefits. 

\begin{figure}[t]
    \centering
    \includegraphics[width=\linewidth]{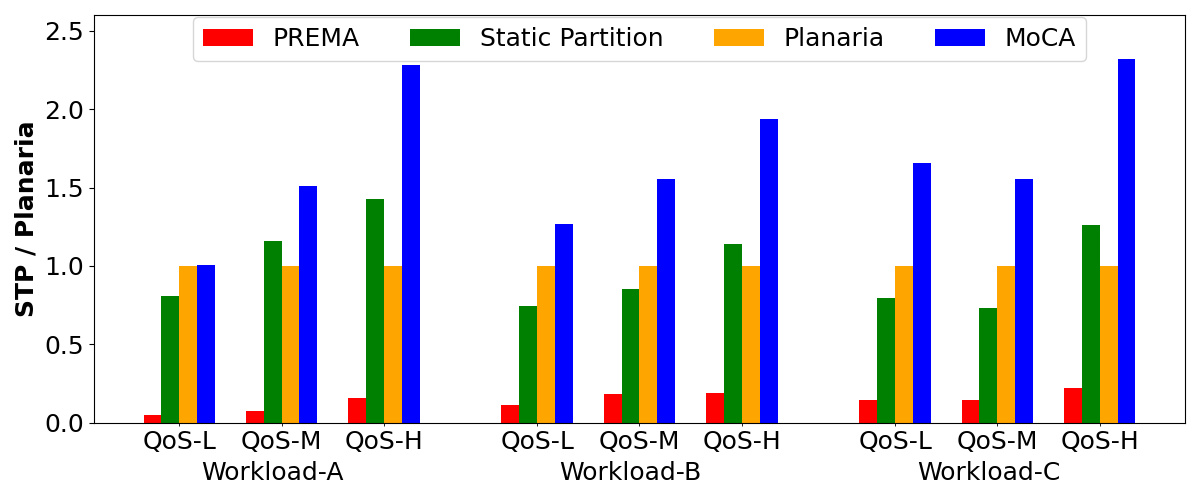}
    \caption{STP improvement of \sys over evaluated multi-tenancy baselines (normalized to Planaria baseline) with different QoS targets and different DNN workload sizes.}
    \label{fig:stp}
\end{figure}

\subsection{Throughput analysis}

Figure~\ref{fig:stp} shows the system throughput improvement that \sys achieves over the baseline multi-tenancy strategies, normalized to Planaria's STP.
\sys achieves a 12.5$\times$ geometric mean improvement over Prema, and up to a 20.5$\times$ maximum improvement (for Workload-A, QoS-L and Workload-A, QoS-M).
Compared to the static partitioning baseline, \sys achieves a 1.7$\times$ higher geometric mean, with up to 2.1$\times$ improvement in certain workloads (Workload-C, QoS-M).
Finally, \sys achieves a 1.7$\times$ geometric mean improvement over Planaria, with a 2.3$\times$ maximum improvement (Workload-C, QoS-H and Workload-A, QoS-H).

Workload-A (w/ light models) shows the most improvement on \sys compared to Planaria, with most of this improvement resulting from \sys's less frequent threading overhead.
Planaria shows poor throughput for the QoS-H scenario on lighter DNN models, as it is unable to utilize its compute resources effectively during thread migrations, whose overhead is comparable to the actual runtime of the lighter models.

For the Workload-B group (with heavy models), \sys adaptively modulates the contention between heavy DNN models, leading to an overall improvement in STP.
The \sw dynamically detects contention during runtime, and the \hw can then resolve contention by throttling excessive memory accesses from memory-bounded layers up to a limit, calculated based on priority and slack time.
With this control over the memory access rate, \sys can facilitate the progress of co-running applications better than the baseline alternatives, increasing the overall STP.

Workload-C (with all models) shows further improvements, mainly because it included a better mix of memory- and compute-bound co-located DNN layers.
Layer grouping between memory-intensive layer groups and compute-bound layer groups later in the task scheduling queue helps increase PE array utilization, resulting in higher system throughput.
Although Planaria does compute resource repartitioning, its effect is limited when the performance degradation originates from memory contention, since its scheduler does not consider the memory requirements of co-running applications.
In this case, multiple memory-intensive layers can be scheduled concurrently, leading to suboptimal performance. 


\begin{figure}[t]
    \centering
    \includegraphics[width=\linewidth]{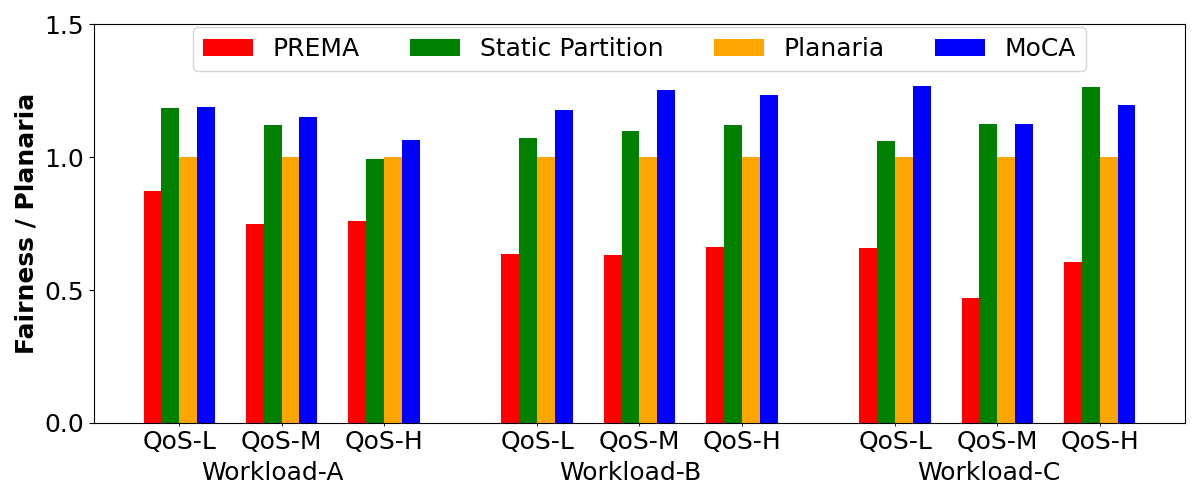}
    \caption{Fairness improvement of \sys over evaluated multi-tenancy baselines (normalized to Planaria baseline) with different QoS targets and different DNN workload sizes.}
    \label{fig:fairness}
\end{figure}

\subsection{Fairness analysis}
We evaluate the fairness of colocation, as defined in Section~\ref{methodology}, to show that \sys improves the throughput at the system level without harming the overall fairness of the system.
Figure~\ref{fig:fairness} compares the fairness of \sys and the three other baseline systems.
Compared to Prema, \sys shows a geometric mean improvement of 1.8$\times$, up to a maximum 2.4$\times$ improvement (Workload-C, QoS-M).
\sys shows a 1.07$\times$ geometric mean improvement over to the static partitioning baseline, and 1.2$\times$ (Workload-C, QoS-L) improvement at most.
Finally, compared to Planaria, \sys shows a 1.2$\times$ geometric mean improvement while 1.3$\times$ (Workload-C, QoS-L) maximum improvement.
In general, \sys shows improved fairness compared to all baselines.
Furthermore, the fairness benefit of \sys was most pronounced for Workload-B, where memory intensive layers are populated as highest rate.
\sys's contention monitoring and resolving mechanism help relieve system level contention so that co-running applications of the memory bounded layers not to be unequally starved.
With memory modulation, \sys fairness outweighs static partitioning counterparts, which does not exploit contention management.

We also observe that \sys delivers slightly lower fairness compared to the static partition system in the case for Workload-C with all the DNN networks. 
This is due to the fact that when there are a diverse set of layers available with different compute and memory requirements, \sys's memory-aware scheduler tends to group short-running, compute-intensive layers together with long-running, memory-intensive layers to maximize the system-level throughput with a balanced workload combination, as demonstrated in Figure~\ref{fig:stp}.
At the same time, this may make some of the lower-priority, short-running models finish earlier than needed, hence, lower the fairness metric slightly.

\begin{figure}[t]
    \centering
    \includegraphics[width=1.0\linewidth]{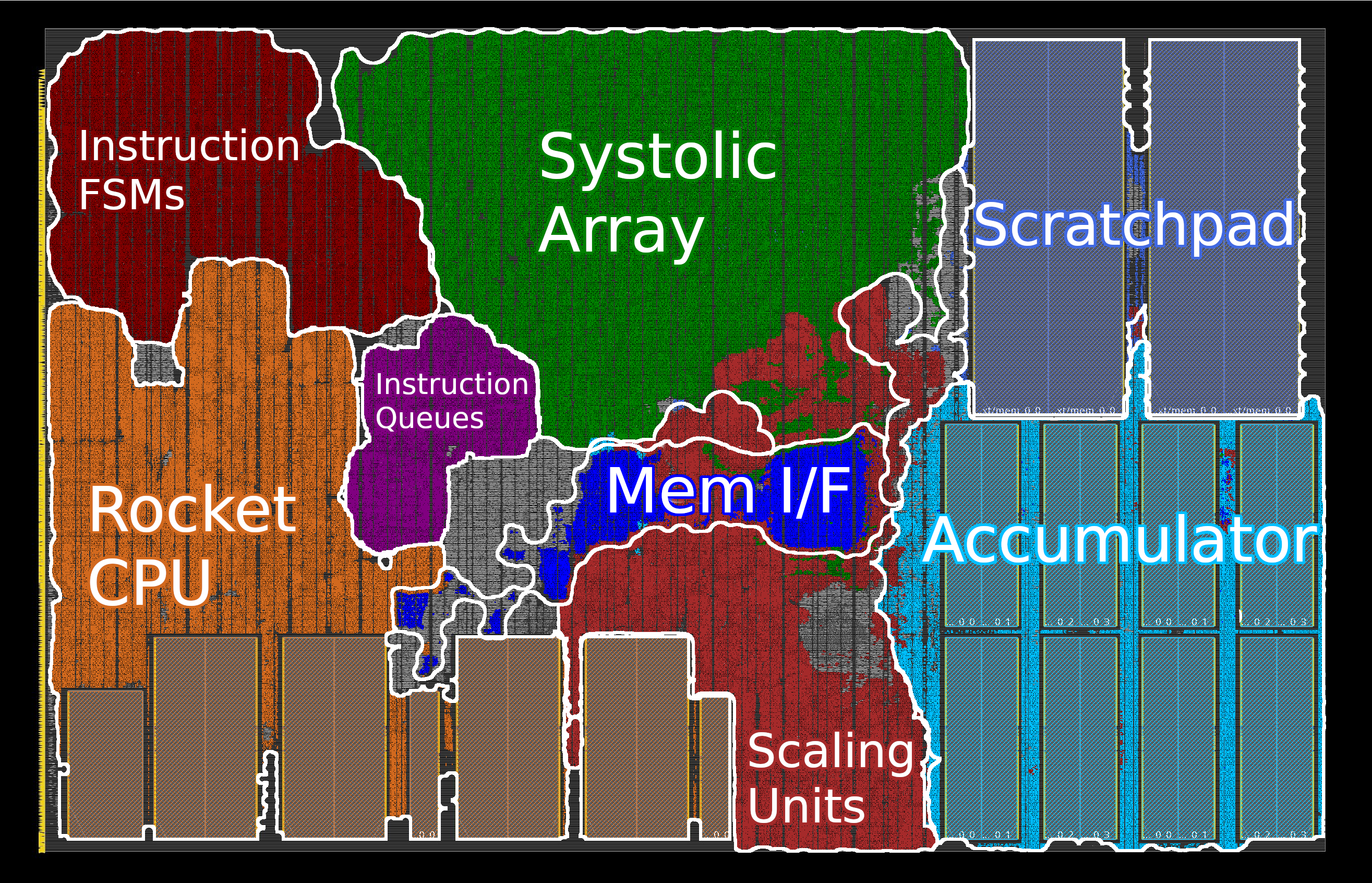}
    \caption{Layout of an accelerator tile with \sys.}
    \label{fig:pnr}
\end{figure}

\begin{table}[t]
\centering
\scalebox{1.1} {
\begin{tabular}{ l | r | r }
\hline
\textbf{\makecell{Component}} &  \textbf{\makecell{Area ($\mu \mathrm{m ^2}$)}} & \textbf{\makecell{\% of System Area}} \\
\hline
\hline
\makecell{Rocket CPU} & 101K & 20.5\% \\ \hline
\makecell{Scratchpad} & 58K & 11.7\% \\ \hline
\makecell{Accumulator} & 75K & 15.2\% \\ \hline
\makecell{Systolic Array} & 78K & 15.8\% \\ \hline
\makecell{Instruction Queues} & 14K & 2.8\% \\ \hline
\makecell{Memory Interface\\w/o \sys} & 8.6K & 1.7\%  \\ \hline
\makecell{\hw} & 0.1K & 0.02\%  \\ \hline
\makecell{Tile} & 493K & 100\% \\ \hline
\hline
\end{tabular}
}
\caption{Area breakdown of an accelerator tile with \sys.}
\label{tab:area-table}
\end{table}

\subsection{Physical design and area analysis}

We synthesized a \sys-enabled DNN accelerator using Cadence Genus with GlobalFoundries' 12nm process technology.
We also place-and-routed the \sys-enabled DNN accelerator using Cadence Innovus, as shown in Figure~\ref{fig:pnr}.
The \hw is implemented in the memory-interface of the accelerator.

As shown in Table~\ref{tab:area-table}, \sys adds only a small area overhead to the DNN accelerator, increasing the size of the accelerator's memory interface by just 1.7\%, and the area of the entire accelerator (including functional units) by 0.02\%.

}

\section{Conclusion}\label{conclusion}
{\rm
An adaptive, memory-centric multi-tenancy accelerator architecture for DNN workloads, \sys, has been proposed, implemented and analyzed in this paper.
Differing from the existing solutions that aim to partition the compute resources of DNN hardware, \sys focuses on dynamically managing the shared memory resources of colocated applications to ensure satisfying QoS targets while considering priority difference. 
\sys leverages the regularity of DNNs and accelerators to estimate and manipulate the usage of memory resources based on applications' latency targets and user-defined priorities.
Our thorough evaluation on a diverse set of DNN execution scenarios and varied target constraints demonstrates that \sys can increase the SLA satisfaction rate up to 3.9$\times$, with a geometric mean speedup of 1.8$\times$ for overall workloads, as well as up to 4.7$\times$ for high priority workloads over prior works. 
\sys also increases the overall system throughput up to 2.3$\times$ while also improving fairness up to 1.3$\times$.
Finally, \sys provides these benefits while incurring less than 1\% area overhead of a state-of-the-art DNN accelerator.

}
\section{Acknowledgements}

This research was, in part, funded by the U.S. Government under the DARPA RTML program (contract FA8650-20-2-7006). 
This work was also supported in part by the NSF Award CCF-1955450 and in part by SLICE Lab industrial sponsors and affiliates.
The views and conclusions contained in this document are those of the authors and should not be interpreted as representing the official policies, either expressed or implied, of the U.S. Government. 


\bibliographystyle{IEEEtranS}
\bibliography{seah,sophia-v2}

\clearpage
\appendix

\newcommand{\mytextbox}[2][6]{
\begin{framed} \small
\noindent#2%
\multido{\i=1+1}{#1}{\newline …}
\end{framed}}

\thispagestyle{empty}

\setcounter{section}{0}



\subsection{Abstract}
This artifact appendix section describes how to access the artifacts for each \sys component, exercise and evaluate it as in Section~\ref{result}.
As in Section~\ref{methodology}, we will use FireSim FPGA-accelerated simulations to evaluate \sys in a full-stack environment.

\subsection{Artifact meta-information checklist}
{\small
\begin{itemize}
    \item \bf Runtime environment: AWS FPGA Developer AMI 1.12.1
    \item \bf Hardware: AWS EC2 instances (c5.4xlarge, f1.2xlarge)
    \item \bf How much disk space is required?: 200 GB (on EC2 instance).
    \item \bf Experiments: FireSim simulations of \sys incorporated into a RISC-V-based DNN accelerator infrastructure, running multi-tenant inference queries.
    \item \bf Program: Chisel (RTL), C (Runtime, Scheduler), Python (Script)
    \item \bf Metric: Target satisfaction rate (rate of each query that meets the target deadline), STP (System Throughput), and Fairness as defined in Section~\ref{subsec:metrics}.
    \item \bf Output: Parsed result from UART output of SoC, performance (SLA satisfaction rate, STP, Fairness) comparison bar plot between static partitioning baseline and \sys for each QoS level and Workload sets. (Figure 5-8)
    \item \bf How much time is needed to prepare the workflow?: 2 hours (scripted installation).
    \item \bf How much time is needed to complete experiments?: 4 hours (scripted run, scripted result parsing)
    \item \bf Publicly available: Yes.
    \item Code licenses: Several, see download.
\end{itemize}
}

\subsection{Description}
\subsubsection{How to access}
The artifacts consist of:
\begin{enumerate}
    \item FireSim: Top-level FPGA-Accelerated RTL Simulation Environment (\url{https://doi.org/10.5281/zenodo.7456139}) 
    \item Chipyard: RISC-V SoC generation environment (\url{https://doi.org/10.5281/zenodo.7456073}) 
    \item \sys Hardware: \sys implementation on Gemmini DNN accelerator (\url{https://doi.org/10.5281/zenodo.7456052}). The main implementation is on DMA.
    \item \sys Software: \sys runtime and scheduler implementation, and tests. (\url{https://doi.org/10.5281/zenodo.7456045}). 
    \sys runtime and schedulers are under \texttt{imagenet} and \texttt{include}.
\end{enumerate}

Users need not download the latter three repositories manually—they will be obtained automatically from Zenodo when the FireSim repository is set up.

\subsubsection{Dependencies - Hardware} One AWS EC2 c5.4xlarge instance (also referred to as “manager” instance), and three f1.2xlarge instances are required (we split the workload to run on three parallel f1 instances to save runtime, which takes more than 8 hours if run on a single instance). The latter will
be launched automatically by FireSim’s manager.
We have provided pre-built FPGA images to avoid the long latency ($\sim$8 hours) of the FPGA-built process. However, if users want to build custom FPGA images, one additional z1d.2xlarge is required.

\subsubsection{Dependencies - Software} Use ssh or mosh on your local machine to remote access EC2 instances. All other requirements are automatically installed by scripts in the following sections.

\subsection{Installation}
First, follow the instructions on the FireSim website\footnote{\url{https://docs.fires.im/en/1.15.1/Initial-Setup/index.html}} to create an EC2 manager instance. You must complete up to and including “Section 1.3.1.2: Key Setup, Part 2”, with the following recommendations on “Section 1.3.1”:
\begin{enumerate}
    \item When instructed to launch c5.4xlarge or z1d.2xlarge, select c5.4xlarge.
    \item When entering the root EBS volume size, using 200GB is sufficient.
\end{enumerate}
Once you have completed up to and including "Section 1.3.1.2 Key setup" in the FireSim docs, you should have a manager instance set up, with an IP address and key. 
Either do ssh or mosh to log in to the instance.
From this point, all commands should be run on the manager instance.

For artifact evaluation, please clone the forked FireSim repository by running the following (else, users can clone FireSim and checkout the \texttt{MOCA} branch in the Gemmini submodule):

For artifact evaluation, begin by downloading the top-level FireSim repository from Zenodo: 

\mytextbox[0]{

\texttt{\$ wget -O firesim-moca-ae.zip https://zenodo.org/
record/7456139/files/firesim-moca-ae.zip}

\texttt{\$ unzip firesim-moca-ae.zip}
}

Next, run the following, which will initialize all dependencies and run the FireSim and Chipyard setup steps (RISC-V toolchain installation, matching host toolchain installation, etc.):
\mytextbox[0]{

\texttt{\$ cd firesim}
 
\texttt{\$ ./first-clone-setup-fast.sh}
}

After the script finishes running, run the following:
\mytextbox[0]{

\texttt{\$ source sourceme-f1-manager.sh}
}

After sourcing, complete the steps in "Section 1.3.3 Completing Setup Using the Manager".
Once these steps have been completed, you are fully ready to evaluate \sys.

\subsection{Experiment Workflow} \label{sec: workflow}

Now that our environment is set up, we will run \sys artifact.
First, we will begin with building the workload for \sys.
When building the workload image, users can either compile the Linux binary themselves or can use provided pre-compiled binary. 

\subsubsection{Building Linux image containing workload}
\begin{enumerate}
    \item On the manager instance, build the FireSim-compatible RISC-V Linux image using a buildroot-based Linux distribution. Follow the instruction on "Section 2.1.1" of FireSim documentation.\footnote{\url{https://docs.fires.im/en/1.15.1/Running-Simulations-Tutorial/Running-a-Single-Node-Simulation.html}}
    \item Run the following to create \texttt{build} directory at \texttt{gemmini-rocc-tests}.
\mytextbox[0]{

\texttt{\$ cd firesim/target\_design/chipyard/generators}

\texttt{\$ cd gemmini/software/gemmini-rocc-tests}
 
\texttt{\$ ./build.sh}
}
    \item (Optional - If the users want to compile the binaries themselves, please skip this process) To test provided pre-compiled binaries, unzip \texttt{imagenet-binary.zip} to \texttt{build/imagenet} by running the following commands. Comment out all the tests under \texttt{test} in \texttt{imagenet/Makefile} to prevent pre-compiled binaries being overwritten. 
\mytextbox[0]{

\texttt{\$ cd build}

\texttt{\$ rm -rf imagenet/}

\texttt{\$ unzip ../imagenet-binariy.zip}

\texttt{\$ cd ../}

}
    \item Run the following command to build the \sys runtime and scheduler which are written in C on a full Linux environment. Running the commands will generate a \texttt{.json} file in \texttt{firesim/deploy/workloads}.
\end{enumerate}
\mytextbox[0]{

\texttt{\$ cd ../}
 
\texttt{\$ ./build-gemmini-workload.sh}
}

\subsubsection{Running FireSim simulation}
\begin{enumerate}
    \item Go to \texttt{firesim/deploy}, and within \texttt{config\_hwdb.yaml}, paste the pre-built FPGA image entry provided in \texttt{built-hwdb-entries/}. Set this in \texttt{default\_hw\_config} in \texttt{config\_runtime.yaml} as well.
    \item For other configurations in \texttt{config\_runtime.yaml}, see "Section 2.1.2. Setting up the manager configuration" of FireSim documentation. \\ The following parameters must be modified:
    \begin{enumerate}
        \item \texttt{workload\_name: gemmini-tests-workload.json}.
        \item Increase the number of f1.2xlarge to boot to 3 (\texttt{f1.2xlarge: 3}), under \texttt{run\_farm\_hosts\_to\_use}.
        \item Replace \texttt{topology: three\_no\_net\_config} under \texttt{target\_config}.
        \item Set \texttt{no\_net\_num\_nodes: 3} to launch three f1 instances running in parallel.
    \end{enumerate}
    \item Run FireSim simulations by launching f1.2xlarge instances. Follow the instructions on "Section 2.1.3 Launching a Simulation!" of the FireSim documentation.
    \item The result will be copied to a directory in \\ \texttt{deploy/results-workload}. The generated result directory will consist of three sub-directories, each for workload-A/B/C defined on Table~\ref{tab:workloads}, as each f1 instance run each workload type. Running the following commands on \texttt{results-workload} directory will generate figures for each workload set. Run the first command from the following command block to parse the results in the priority level group, as well as the general results as Figure~\ref{fig:sla_priority} and Figure~\ref{fig:sla_overall} per each QoS level and workload model set. The second command is to parse the STP and Fairness results as in Figure~\ref{fig:stp} and Figure~\ref{fig:fairness}. The result directory on the script (\texttt{\$result\_dir}) should be the directory that includes three sub-directories. 
\end{enumerate}
\mytextbox[0]{

\texttt{\$ ./build\_sla.sh (\$result\_dir)}

\texttt{\$ ./build\_stp.sh (\$result\_dir)}
}

The test scripts will run the static partition baseline and \sys on 250 end-to-end turns of inference queries that are dispatched randomly, with randomly assigned priority. 
Note that this script will not rebuild FPGA images for the system by default, since each build takes around 8 hours. We instead provide pre-built images by default on \texttt{built-hwdb-entries/}, which is used on the paper's evaluation.

The result parsing scripts that are used in build commands (\texttt{build\_sla.sh} and \texttt{build\_stp.sh}) are \texttt{parse\_result\_from\_uartlog.py} and \texttt{make\_fair.py}, respectively. Those Python scripts and build scripts are included in \texttt{results-workload} directory. 

We provide example result plots of the pre-compiled binaries under \texttt{EXAMPLE\_RESULT/} for reference.

Please make sure the running f1 instance is terminated by running \texttt{firesim terminaterunfarm}, and confirm in your AWS EC2 management console that no instances remain beside the manager.

\subsection{Experiment customization}

\subsubsection{Rebuilding FPGA image}
Users can change the SoC configuration by changing Gemmini DNN accelerator configuration or other SoC configuration.
For example, on \texttt{Config.scala} of Gemmini \texttt{src}, users can reconfigure the internal scratchpad or accumulator size.
The shared L2 size can be changed by modifying cache parameters at \texttt{RocketConfigs.scala}. 
For building a new FPGA bitstream, please follow the steps in "Section 3. Building Your Own Hardware Designs"\footnote{\url{https://docs.fires.im/en/1.15.1/Building-a-FireSim-AFI.html}}.

\subsubsection{Customizing experiment parameters}
Go to \texttt{gemmini/software/gemmini-rocc-tests/include}. 
In the directory, header files are included for \sys runtime, software, and parameters.
Open \texttt{gemmini.h}, and change \texttt{SEED} to change the seed to generate new randomized queries.
To change the number of queries, change \texttt{total\_workloads}.
Repeat Section~\ref{sec: workflow} to get the result.

\thispagestyle{empty}
\end{document}